\documentclass[11pt]{article}
\usepackage{array}
\usepackage{graphicx}
\usepackage{cite}
\usepackage{textpos}

\title{Physics in 100 Years}

\author{Frank Wilczek}

\begin{document}

\maketitle


\begin{textblock*}{5cm}(11cm,-8.2cm)
  \fbox{\footnotesize MIT-CTP-4654}
\end{textblock*}

\begin{abstract}
Here I indulge in wide-ranging speculations on the shape of physics, and technology closely related to physics, over the next one hundred years.   Themes include the many faces of unification, the re-imagining of quantum theory, and new forms of engineering on small, intermediate, and large scales.
\end{abstract}

\medskip

When Leon Cooper, on behalf of Brown University, asked me to contribute to their 250$^{\it th}$ anniversary by giving a talk about the next 250 years of physics, I of course accepted immediately.  

Then I thought about it.  I soon realized that I'd taken on a task that is way beyond me, or (I suspect) anyone else.   The most impressive attempt of this kind I know of is the set of  {\it Queries\/} that Isaac Newton attached to his {\it Opticks\/} in 1718.   The {\it Queries\/} contain many important ideas and suggestions, but by 1818, after 100 years, they were showing their age\footnote{Unlike Brown University, which is self-renewing.}.   By that  time Lavoisier's scientific chemistry was already well along, the wave theory of light was being established, Oersted and Faraday were just over the horizon, and most of the {\it Queries\/} had been answered, or transcended.   So as a first step I renormalized 250 $\rightarrow$ 100.  

On further reflection, I also came to realize that the point of this talk should not be to predict future history accurately, as in a test or a business plan.   I interpret my assignment, rather, as an opportunity to exercise disciplined imagination.   For the exercise to have intellectual value, beyond mere entertainment or hype, it should be grounded in judgements of three kinds
\begin{itemize}
\item What are the weak points in our current understanding and practices?
\item What are the growth areas in technique and capability?
\item Where are the sweet spots, where those two meet?
\end{itemize}

Here, in that spirit, I've assembled a baker's dozen of Query-like mini-inquiries.    They are loosely structured: Some overlap with others, and some range over several sub-topics.   Here and there, by way of summary, I will venture a few provocative predictions.  Those are italicized below.

\bigskip

\section{A Diversity of Unifications}

Our first seven items concern different sorts of unification.

On one occasion, following a public lecture about the possibility of a unified theory of the four forces, an audience member posed me this disarming question:
\begin{quote}
Why do physicists care so much about unification?  Isn't it sufficient just to understand things?
\end{quote}
It's a profound question, actually.    It is not at all self-evident, and may not be true, that going after unification is an efficient way to advance knowledge, let alone to address social needs or to promote human welfare.   I can offer two answers:

\begin{itemize}
\item{\it Historical}: It is apparent, in retrospect, that many of the greatest advances from the past have been unifications of disparate subjects.   In most of these cases, when the unification took place each side of the separate subjects was already well developed on its own, and the ``value added'' by unification was not immediately obvious.  But in every case the unification itself proved extremely fertile, in unanticipated directions:
\begin{itemize}
\item Unification of algebra and geometry (Descartes)\footnote{I've attached names to the developments not to assign credit, but as shorthand to recall some central contributions.}
\item Unification of celestial and terrestrial physics (Galileo, Newton)
\item Unification of mechanics and optics (Hamilton)
\item Unification of electricity, magnetism, and optics (Maxwell)
\item Unification of space and time (Einstein, Minkowski)
\item Unification of wave and particle (Einstein, de Broglie)
\item Unification of reasoning and calculation (Boole, Turing)
\end{itemize}

\item{\it Esthetic}: Isn't it pretty to think so?

\end{itemize}

\bigskip

\subsection{Unification I: Forces}

The Core Theory of physics, which currently provides our best description of physical fundamentals, became firmly established in the later part of the twentieth century.  It consists of smoothly connected set of quantum field theories, all based on local symmetry, that describe the strong, weak, electromagnetic and gravitational interactions\footnote{The term ``Standard Model'' has many shortcomings, not the least of which is that it is sometimes used for the electroweak theory, sometimes specifically for the minimal electroweak model, sometimes for the electroweak theory together with quantum chromodynamics, and sometimes for the theories of all four forces.   For reasons I've detailed in an Appendix, I think the most sensible procedure is to use ``Standard Model'' in its original sense, to mean the electroweak theory only.  I also think that it is perverse to exclude gravity, in the form of quantized general relativity, from the inventory of our established fundamental understanding.   A different term is needed for the whole shebang, and here I will use Core theory, or when appropriate Core theories.}   We may summarize the theory through its local symmetry group
$$
SU(3) \times SU(2) \times U(1) \times SO(3,1)
$$

\begin{figure}[h!]
\begin{center}
\includegraphics[scale=.35]{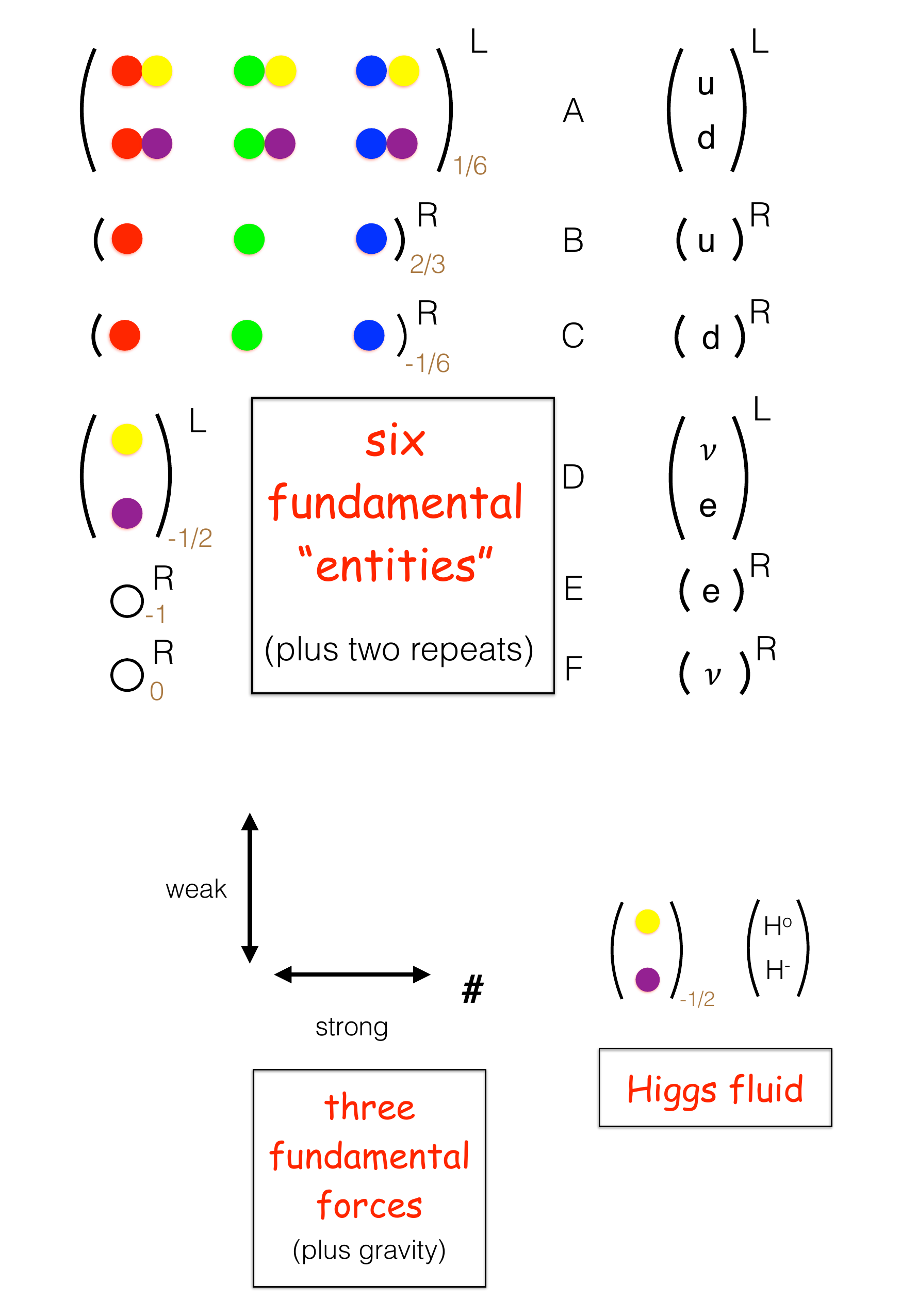}
\caption{Because the physical implications of local symmetry are extremely powerful, we can honestly convey the central characteristics of the strong, weak, and electromagnetic forces in the form of a colorful cartoon.  The strong force, based on local $SU(3)$, is a theory of the transformations among three color charges, here denoted red, green, and blue; the weak force, based on local $SU(2)$, is a theory of transformations among two color charges, here denoted yellow and purple; the electromagnetic force (properly: the hypercharge force), based on local $U(1)$, is a theory of complex number phase transformations.  The nonabelian $SU(3)$ and $SU(2)$ representations relevant to Core substance are the simplest non-trivial ones, i.e. the basic vector representations, while the $U(1)$ representations bring in numbers, here displayed as subscripts. The substance (fermion) particles within each family of the Core break up into six irreducible entities.  Their conventional names are shown on the right.  The triplication of families, and especially the couplings of the Higgs field,  bring in complications -- masses and weak mixing angles -- that local symmetry only loosely constrains.}
\label{coreTheory}
\end{center}
\end{figure}

\begin{figure}[h!]
\begin{center}
\includegraphics[scale=.35]{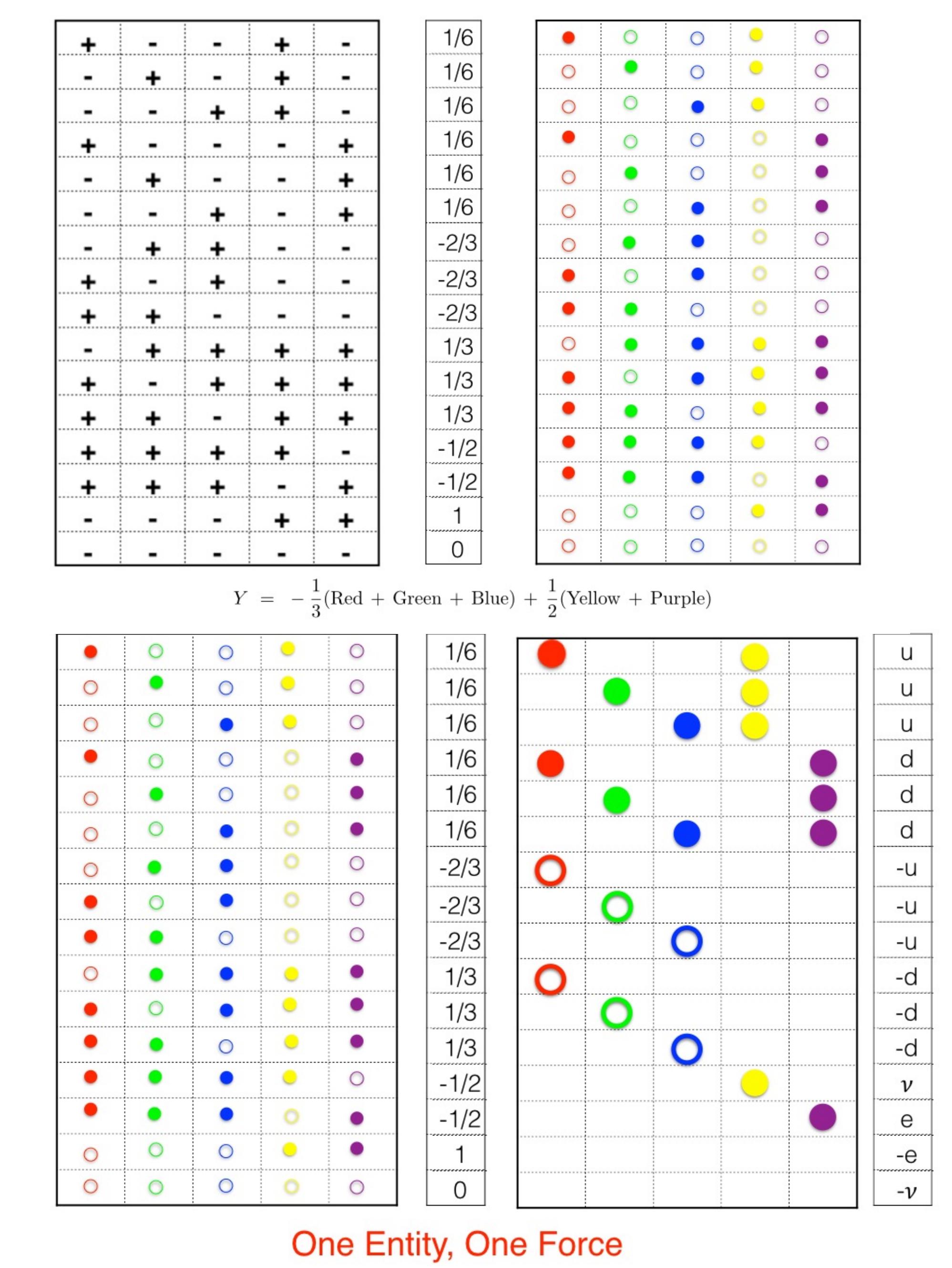}
\caption{The local symmetries of the Core can be accommodated beautifully into an encompassing unified symmetry.  The most straightforward and (to me) compelling version of this idea is based on the symmetry $SO(10)$ and the 16-dimensional spinor representation for substance, as first proposed by Georgi and Glashow.  For further explanation, see the text.}\label{so10Unification}
\end{center}
\end{figure}

We can summarize the most striking {\it qualitative\/} indication for unification of forces in a pair of iconic images (Figures \ref{coreTheory}, \ref{so10Unification}).

\pagebreak

\pagebreak

\pagebreak

At the level of color charges -- i.e., according to the $SO(2)\times SO(2) \times SO(2) \times SO(2) \times SO(2)$ Cartan subgroup, which serves here as a complete set of commuting observables --  we find all possible combinations of half-units of positive and negative charge.  (The charges for each color are displayed in the columns, and the rows represent different particles.)   This is shown abstractly on the upper left, and in appropriate color notation on the upper right.  Hypercharge assignments are no longer an independent choices, but is related to the strong and weak color charges according to the middle formula, whose results for each row are tabulated in the central column.   In the bottom part of this Figure we first reproduce, and then process, part of the top.   Our processing uses the strong and weak ``bleaching rules'' to simplify the depiction of strong and weak color charges.  According to the strong bleaching rule we can add equal amounts of red, green, and blue color charge, without changing the strong interaction; the weak bleaching rule is similar.   With tasteful application of these rules, we arrive at the version of the table you see on the lower right.  As a last step, we recognize that left-handed particles, which is what the construction uses, will also serve to represent their right-handed antiparticles, after reversing the signs of all the charges.   With that, we discover that a perfect match between the properties of names in the final column and the corresponding substance particles of the Core!

A special advantage of this scheme is that it makes the appearance of small but non-zero neutrino masses natural.  This feature was appreciated, and advertised, well before the experimental observation of such masses.    Thus it is fair to regard those observations as semi-quantitative validation of the unification framework\footnote{For experts: Note that the existence of the $SU(3)\times SU(2) \times U(1)$ singlet ``right handed neutrino'', which is crucial here, is {\it not\/} required for anomaly cancellation.}.

\bigskip

Another striking, though not yet fully realized, success of the unification idea is its explanation of the relative strengths of the different forces.  I will discuss the quantitative aspects of unification further below, in Unification II.    

These successes: ``derivation'' of the scattered multiplet structure of $SU(3)\times SU(2) \times U(1)$ fermions from a beautiful, irreducible representation; prediction of small but non-zero neutrino masses; and (contingently) quantitative explanation of the relative strengths of the different forces -- are, I think, extremely impressive.    It is difficult to believe that they are accidental.   

This unification of forces, however, raises several challenges.  

On the experimental side:  Besides non-zero neutrino masses, the other classic experimental implication of unification is proton instability, with a very long but perhaps not inaccessible lifetime.  That prediction has not yet been verified, despite heroic efforts.   The existing limits put significant pressure on the framework.  Reading it optimistically: There is an excellent chance that further efforts along this line would be rewarded.
 
On the theoretical side: While the pattern of gauge bosons and fermions fits expectations from unified theories beautifully, the Higgs doublet does not.   Also, so far attempts to understand the origin of family replication and the detailed pattern of fermion masses and mixings based on unification ideas have been inconclusive at best.   I'm not confident that the situation will be markedly better in 100 years, though I'd love to be proved wrong (and even more, I'd love to {\it see myself\/} proved wrong).   The only reasonably compelling idea in this area, in my opinion, is the Peccei-Quinn mechanism for understanding approximate $P$ and $T$ symmetry.   It leads one to expect the existence of axions -- an expectation with major cosmological implications.    

With that background, I will hazard two predictions for physics in the next 100 years:
\begin{itemize}
\item {\it Proton decay will be observed.  Study of baryon number violating decay modes will become a rich and fruitful field of empirical science.}
\item {\it Axions will be observed, as a cosmic background.  The study of that background will become a rich and fruitful branch of cosmology and fundamental physics.}
\end{itemize}

\bigskip

\subsection{Unification II: Force and Substance}

\bigskip

\begin{figure}[h!]
\begin{center}
\includegraphics[scale=.7]{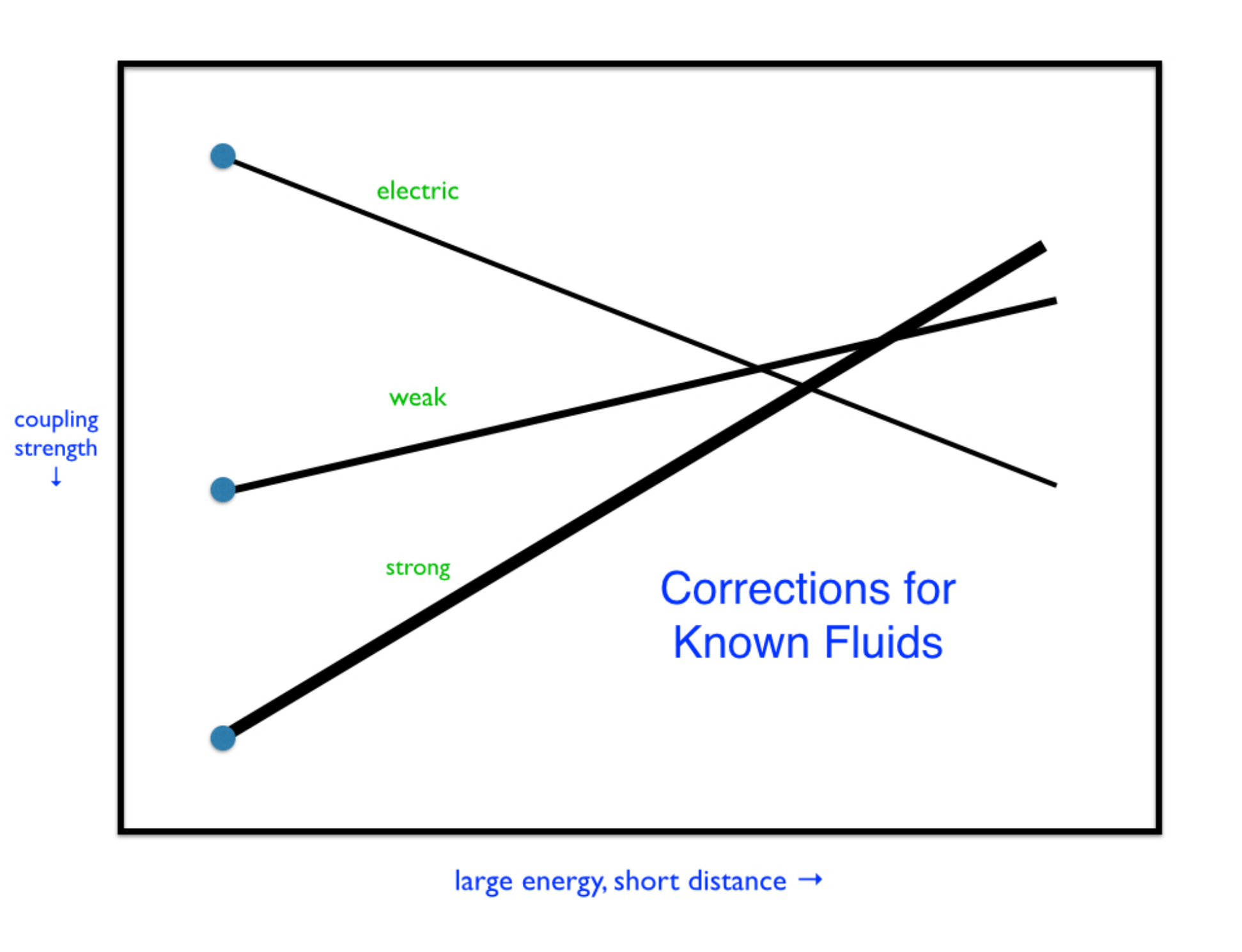}
\caption{If we extrapolate the observed couplings up to high energy, including the known particles within our renormalization group equations, we find that they nearly unify, but not quite.}
\label{couplings}
\end{center}
\end{figure}

\begin{figure}[h!]
\begin{center}
\includegraphics[scale=.7]{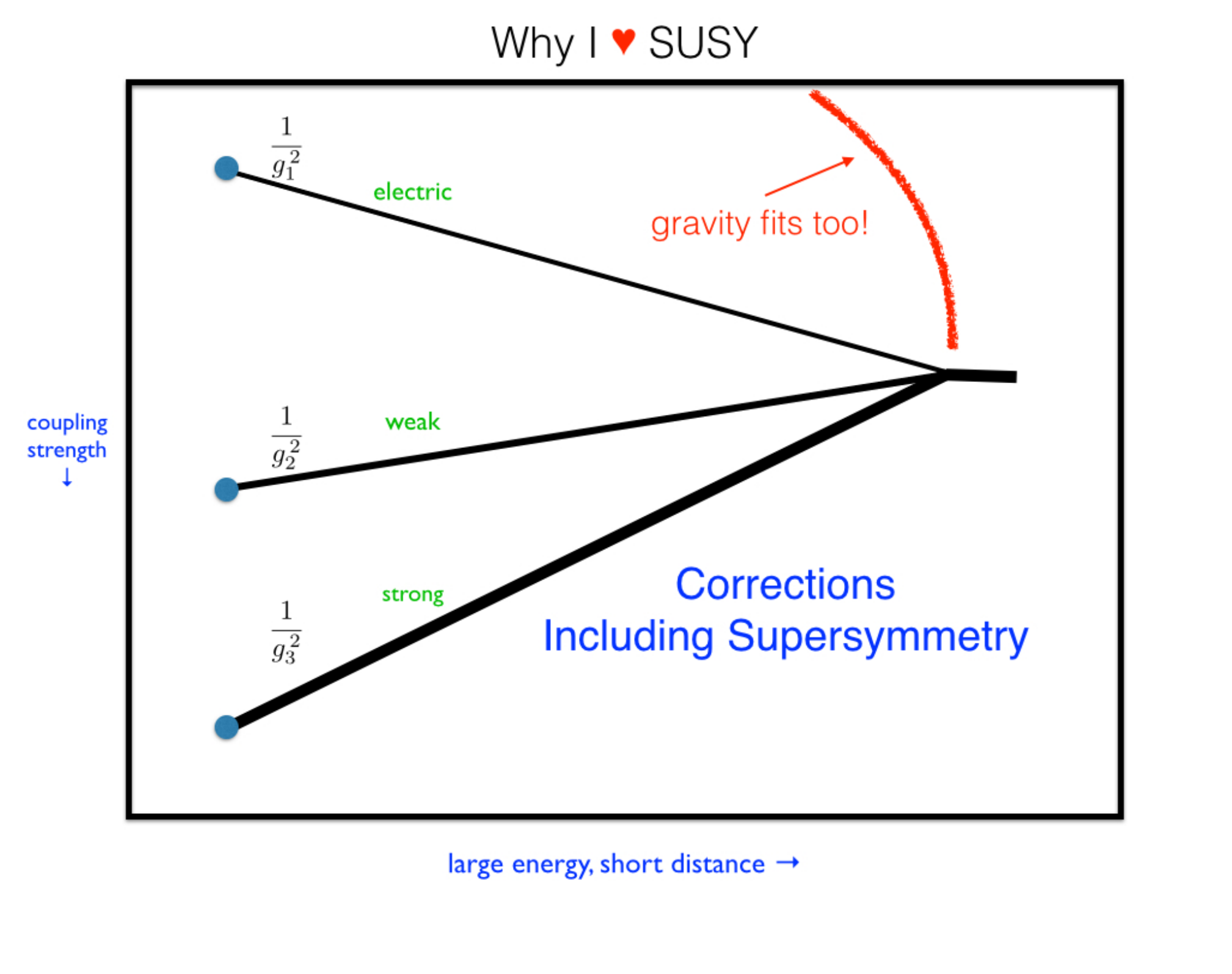}
\caption{If we include, in addition, the effects of the hypothetical particles needed to accommodate supersymmetry, we get accurate unification of the three gauge forces, and semi-quantitative unification with the gravitational force.}
\label{susyUnification}
\end{center}
\end{figure}

We can summarize the most striking {\it quantitative\/} indication for unification of forces in a second pair of iconic images (Figures \ref{couplings}, \ref{susyUnification}.)   

\pagebreak

\pagebreak

\pagebreak

The unification of quantum numbers, as indicated in Figure \ref{so10Unification}, is not in itself a fully formed theory.   We need to give an account of how the ``un-unification'' took place.  This can be done most straightforwardly using a generalization of the Higgs mechanism (now triumphant in its explanation of a more modest symmetry breaking, that of electroweak $SU(2)\times U(1)_Y$ to $U(1)_Q$).   

The un-unification is associated with a characteristic energy scale, the magnitude of the symmetry breaking condensate.    If we evolve the gauge coupling constants according to the renormalization group, then they should become equal beyond that scale.   Since we have three independent couplings for $SU(3)\times SU(2) \times U(1)$, but only two parameters -- the scale of the symmetry breaking, and the value of the unified coupling --  that requirement leads to a constraint among observed quantities, that might or might not hold in Nature. 

The evolution of couplings according to the renormalization group depends, of course, upon the spectrum of particles that contribute.   Additional particles whose masses lie between currently observed scales and the unification scale will contribute to vacuum polarization, and alter the evolution.   

If we do the calculation using only the contributions of known particles, the constraint implied by unification does not quite work.  That is shown in Figure \ref{couplings}.   

But there is a specific expansion of the spectrum that is attractive in many ways, which should also be considered.   The unification of forces, even if it were perfected, would leave us with two great kingdoms of particles, still not unified.  Technically, these are the fermion and boson kingdoms.  More poetically, we may call them the kingdoms of substance (fermions) and force (bosons).   

By postulating that the fundamental equations enjoy the property of supersymmetry, we heal the division of particles into separate kingdoms.   Supersymmetry can be approached from several different angles, but perhaps the most appealing is to consider it as an expansion of space-time, to include quantum dimensions.  The defining characteristic of quantum dimensions is that they are represented by coordinates that are Grassmann numbers (i.e., anticommuting numbers) rather than real numbers.    Supersymmetry posits that the fundamental laws of physics remain invariant transformations that correspond to uniform motion in the quantum dimensions.  Thus supersymmetry extends Galileo/Lorentz invariance.   

The new transformations of supersymmetry transform fermions into bosons, and {\it vice versa}, while leaving their various charges (strong, weak, and electromagnetic) unchanged.  They also don't change the mass.  Because of that last feature, supersymmetry cannot be an exact symmetry of the world, for we certainly wouldn't have overlooked bosonic particles with the same charge and mass as electrons.  But we can maintain the fallback position, that supersymmetry is broken spontaneously, like electroweak $SU(2)\times U(1)$ and our hypothetical force unification symmetry.    If the scale of supersymmetry breaking is less than the force unification scale, we should add the effect of the new particles supersymmetry requires into our renormalization group equations.   

If we assume the effective scale of supersymmetry breaking is close to the electroweak scale, i.e. $\sim$ 1 TeV, we find a most striking result: The observed couplings obey the unification constraint.   That is what is shown on the right in Figure \ref{susyUnification}. 

\bigskip   

The unification mass scale indicated by this calculation is about $2 \times 10^{16}$ GeV.    That is also an encouraging result, for several reasons.   It is large enough, that proton decay through gauge boson exchange is not predicted to be too large, but it is small enough that gravitational corrections to coupling constant evolution (which are very uncertain) are plausibly negligible.  It also accounts reasonably well for the observed value of neutrino masses, through the see-saw mechanism.    

Most profoundly, it allows us to bring gravity into the unification, at least semi-quantitatively.   For the gauge interactions we have a definite quantized unit of charge, which provides a firm starting point for quantitative comparison.   For gravity, the logic is different.  In comparing the power of the forces at different scales, we are using probes with at different energy-momentum transfers.  When energy-momentum transfers get large, they plausibly dominate any contribution of mass to the power of gravitational couplings.  So one may do the calculation by ignoring the mass of our probes; and that is what I've done.    

The direct dependence of the power of gravity on energy-momentum is much steeper than the logarithmic dependence we find for gauge couplings.    That steep dependence lies behind the amazing fact that at the unification scale the power of gravity closely approaches that of the other interactions.    Let me remind you, that at accessible energies the gravitational interaction among elementary particles is {\it absurdly\/} smaller than the other interactions, by factors $< 10^{-40}$!    

\bigskip

The successful unification of couplings calculation I have reviewed uses a minimal implementation of supersymmetry and a breaking scale $\sim$ 1 TeV.  In other words, we construct the smallest supersymmetric theory that includes the particles we know, and assign mass 1 TeV to all those that haven't been observed yet.    Of course, if the result depended sensitively on those specific assumptions it would be of very limited value.    Fortunately those assumptions can be relaxed, in two distinct ways, without upsetting the good consequences we've mentioned above.    First:  Neither the fact of unification, nor the scale at which it occurs, is much affected by the addition or subtraction of complete unified $SO(10)$ (or $SU(5)$) multiplets.   In particular, we can add any number of singlet ``hidden sector'' fields, or extra fermion families; and we can raise the masses of the squarks and sleptons (but not the gauge partners) freely, in effect suppressing their contribution.    Second: Since the evolution of gauge couplings is logarithmic, our calculation's dependence on the scale of supersymmetry breaking is modest.   That scale might be raised to a few 10s of TeV, before we run into serious trouble.  Also, of course, the masses of various supersymmetric partners can differ from one another significantly, so long as their effective average lies in this broad range.   

\bigskip

It is difficult to believe that all this qualitative and quantitative success is entirely coincidental.  Accordingly:

\begin{itemize}
\item {\it Supersymmetric partners will be observed.  Their study will open up a new golden age for particle physics.}
\end{itemize}

\bigskip

\subsection{Unification III: Space-Time and Matter}

\bigskip

Although general relativity is based on broadly the same principle of local symmetry that guides us to the other interactions, its implementation of that principle is significantly different.   The near-equality of unified coupling strengths, as we just discussed, powerfully suggests that there {\it should be\/} a unified theory including all four forces, but that fact in itself does not tell us how to achieve it.  

String theory may offer a framework in which such four-force unification can be achieved.   This is not the event, and I am not the person, to review the vast amount of work that has been done in that direction, so far with inconclusive results.   It would be disappointing if string theory does not, in future years, make more direct contact with empirical reality.   There are many possibilities, including some hint of additional spatial dimensions (i.e., a useful larger broken symmetry $SO(1, N) \rightarrow SO(1,3)$), cosmic strings with unusual properties, or explanation of some existing relationships, for example by supplying constraints among the ``free'' parameters in the Core (of which there are many, both continuous and discrete)\footnote{To my mind the existence of gravity, sometimes mentioned in this connection, does not count as a prediction of string theory.  Sorry, but it's been a long time since people thought that things fall sideways.  Also: Deep knowledge of the ideas behind general relativity go into formulating string theory, so it's not an endogenous output.}.    Supersymmetry is an important ingredient in many string theory constructions, so in a generous spirit one could include it in this list; but supersymmetry can also stand on its own, and as an historical matter it was introduced independently.

At present there is not much data for a four-force unification, or quantum gravity in general, to explain!  Perhaps relic gravity waves from inflation or some subtle physical effect that has so far escaped notice will open a window.   Because the minimally coupled Einstein theory has only two accessible parameters -- namely, Newton's constant and the cosmological term --  at the level of basic interactions there is not much data to work with.    Many models for breaking supersymmetry invoke gravitational interactions in a central way (``gravity mediation''), so if supersymmetric particles are discovered they might give us more.   

Optimistically:

\begin{itemize}
\item {\it The unification of gravity with the other forces will become more intimate, and have observable consequences.}
\end{itemize}

We can look forward with confidence to an ``operational'' intermingling of matter and space-time, with the development of gravitational wave astronomy:

\begin{itemize}
\item {\it Gravity waves will be observed, and will evolve into a routine tool for astrophysical and cosmological exploration.  Many sources will be identified, and our knowledge of neutron stars and black holes will reach new levels of detail.}
\end{itemize}

\bigskip

The use of real numbers as our basic model of spatial and temporal continua is based on an extrapolation of classical measuring procedures far beyond their historical roots, into the realm of the infinitely small.   So far, it has served physics extremely well, and survived the twentieth-century revolutions of relativity, quantum theory, and local symmetry.    There are, however, beautiful, rigorous ideas that could fill the continuum more amply.    It would be a pity if Nature did not use them.   Accordingly:

\begin{itemize}
\item {\it Infinitesimals, or some other unconventional number system, will play an important role in the description of space-time.}
\end{itemize}


\bigskip

\subsection{Unification IV: Evolution and Origin}

\subsubsection{Working from Initial Conditions}

At present our fundamental laws are dynamical laws, that describe how a given state of the world evolves into others.   They can also, in principle, be used to extrapolate backward in time.   Those procedures of prediction and reconstruction may become impractical, or impossible, for several reasons:
\begin{enumerate}  
\item The equations exhibit extreme sensitivity to small perturbations in initial conditions, and our empirical knowledge of initial conditions is always imperfect in practice.
\item The existence of horizons is a fundamental constraint on the empirical determination of initial conditions. 
\item One can only fully calculate quantum evolution given full knowledge of the wave function, which can not be obtained empirically.
\item When one relates the computed wave function to observed reality,  quantum theory brings in probabilities.  This may or may not be a consequence of the preceding item.
\item The extrapolation may lead to singularities, or -- what amounts to the same thing -- to such extreme conditions that our theories become unreliable.   Notoriously, this actually happens in our present-day working models of physical cosmology, when we extrapolate backwards in time.
\end{enumerate}

A major part of the art of physics, including thermodynamics and statistical mechanics, not to mention the physical branches of engineering, consists in finding ways to work around those limitations.  One identifies concepts and objects that evolve in robust, comprehensible ways.   With the help of computers, this art will no doubt advance dramatically over the next 100 years, as I'll discuss further below.   But sensitivity to small perturbations in initial conditions, horizons, and quantum uncertainty will be finessed, not eliminated.   I don't expect that constraints 1-4 will look {\it fundamentally\/} different in 100 years, or in 250.  

In this context it is worth noting that by adding additional assumptions to our dynamical equations we can extract ``eternal truths'' about the physical world, which are largely free of reference to initial conditions:
\begin{itemize}
\item In dealing with systems that are effectively starved for energy we can get a unique ground state, or something close to it.  Thus we reach the masses of hadrons and atomic nuclei, the periodic table of elements, ... 
\item In dealing with systems that reach a unique thermal equilibrium, or something close to it.  Thus we reach the Hertzsprung-Russell diagram for stars, the standard big bang cosmology, ... 
\end{itemize}

\subsubsection{Are Initial Conditions Necessary?}

Item 5 is of a different nature.  It poses a sharp challenge, and encourages us to revisit an ancient philosophical issue.

\begin{quote}
How could what is perish? How could it have come to be? For if it came into being, it is not; nor is it if ever it is going to be. Thus coming into being is extinguished, and destruction unknown. -- Parmenides
\end{quote}

\begin{quote}
This world, which is the same for all, no one of gods or men has made. But it always was and will be an ever-living fire, with measures of it kindling, and measures going out. -- Heraclitus
\end{quote}

Tension between the ``God's-eye'' view of reality comprehended as a whole (Parmenides, Plato) and the ``ant's eye'' view of human consciousness, which senses a succession of events in time (Heraclitus), is a recurrent theme in natural philosophy.    

Since Newton's time, the Heraclitian view has dominated fundamental physics.   We divide the description of the world into {\it dynamical laws\/} that, paradoxically, live outside of time, and {\it initial conditions\/} upon which those laws act.   The dynamical laws do not determine which initial conditions describe reality.   That division has been enormously useful and successful pragmatically.   On the other hand, it leaves us far short of a full scientific account of the world as we know it.  For the answer 
\begin{quote}
``Things are what they are because they were what they were.''
\end{quote}
begs the question
\begin{quote}
``Why were things that way, and not any other?''
\end{quote}

The division also appears unnatural in the light of relativity theory.  There we learn to consider space-time as an organic whole, with different aspects related by symmetries that are awkward to express if we insist on carving experience into time-slices.   Hermann Weyl expressed this memorably:
\begin{quote}
The objective world simply {\it is}, it does not {\it happen}. Only to the gaze of my consciousness, crawling along the lifeline of my body, does a section of this world come to life as a fleeting image in space which continuously changes in time.
\end{quote}

The most serious attempt to address these issue that I know of is the ``no boundary'' proposal of Hawking and Hartle.   It may well be on the right track, though its reliance on Euclidean quantum gravity comes fraught with technical issues.    In any case it does not quite consummate the sought-for unification because it remains, in essence, a particular choice of initial conditions.   Maybe it is God's choice -- but if so, we'd like to know whether that choice was logically necessary, or is an act of free will!

\begin{itemize}
\item {\it The fundamental laws will no longer admit arbitrary initial conditions, and will not take the form of evolution equations.}
\end{itemize}

\bigskip

\subsection{Unification V: Action and Information}

\bigskip

One unusual measure of progress in physics has been the progression from quantities to their integrals.  Force is the primary dynamical concept in Newtonian mechanics.  In the nineteenth century, with the rise of thermodynamics and statistical mechanics, energy began to feature as a fundamental ingredient.  Quantum mechanics, with the Schr\"odinger equation, accentuated the importance of energy.   But quantum mechanics goes a step further: quantum mechanics, at least as it is conventionally formulated, requires the canonical formalism, which implies an underlying action principle.
In a parallel development, the use of potentials (i.e., integrated fields) in electrodynamics went from being a convenience into a necessity, for potentials appear in the canonical formulation of electrodynamics.

The ``higher'', integrated forms of dynamics are more constrained than the lower, derived forms.  Thus force fields derived from energy principles must be conservative, and dynamical equations that follow from action principles must be capable of being written in canonical, Hamiltonian form.   Two of the Maxwell equations -- the magnetic Gauss law and Faraday's law of induction -- become identities when we introduce potentials.  Is it possible to go further in this direction?

Leaving that interesting question open, let us consider more closely the foundational quantity in present-day physics: action.   Our fundamental laws are most powerfully expressed using Feynman path integrals.  Within that framework action appears directly and prominently, supplying the measure.   And it is at the level of action that our local symmetry principles take a simple form, as statements of invariance.   

Given Planck's constant as a unit, action becomes a purely numerical (dimensionless) quantity.   The world-action is therefore a specific numerical quantity that governs the basic operation of the physical world.   One would like for such a basic quantity to have profound independent meaning.

Information is another dimensionless quantity that plays a large and increasing role in our description of the world.  Many of the terms that arise naturally in discussions of information have a distinctly physical character.  For example we commonly speak of density of information and flow of information.   Going deeper, we find
far-reaching analogies between information and (negative) entropy, as noted already in Shannon's original work.  Nowadays many discussions of the microphysical origin of entropy, and of foundations of statistical mechanics in general, start from discussions of information and ignorance.   I think it is fair to say that there has been a unification fusing the {\it physical\/} quantity (negative) entropy and the {\it conceptual\/} quantity information.

A strong formal connection between entropy and action arises through the Euclidean, imaginary-time path integral formulation of partition functions.   Indeed, in that framework the expectation value of the Euclideanized action essentially {\it is\/} the entropy.   The identification of entropy with Euclideanized action has been used, among other things, to motivate an algebraically simple -- but deeply mysterious -- ``derivation'' of black hole entropy.    If one could motivate the imaginary-time path integral directly and insightfully, rather than indirectly through the apparatus of energy eigenvalues, Boltzmann factors, and so forth, then one would have progressed toward this general prediction of unification:

\begin{itemize}
\item {\it Fundamental action principles, and thus the laws of physics, will be re-interpreted as statements about information and its transformations.}
\end{itemize}

\bigskip

\subsection{Unification VI: Quantum and Symmetry}

\bigskip

Quantum mechanics, like classical mechanics, is more a framework for theory-building than a specific theory.   In using those frameworks to describe Nature, we must add strong hypotheses about the substances and forces under consideration.  

With the emergence of our Core theories, even in their present imperfect form, we have learned that {\it symmetry}, especially in its local form, is a dominant feature of fundamental law.   In our Core theories, the main ``strong hypotheses'' we add to the general framework of quantum mechanics, in order to get a good description of matter, are hypotheses about symmetries and their representations.  Allowing a bit of poetic license, we could say that those essential hypotheses about matter are commutation relations among symmetry generators, in the form
\begin{equation}\label{symmetryRelations}
[ T^a, T^b ] ~=~ i \, f^{ab}_c \, T^c
\end{equation}

Returning to quantum mechanics, and again allowing poetic license, we could say that its central principle is the non-commutativity of conjugate variables:
\begin{equation}\label{qmCommutators}
[ q^a, p_b ] ~=~ i  \, \delta^a_b
\end{equation}

Now of course the resemblance between Eqn.\,(\ref{symmetryRelations}) and Eqn.\,(\ref{qmCommutators}) leaps to the eye.  This emphases two things:
\begin{itemize}
\item It would be desirable to have quantum mechanics itself formulated as a symmetry principle.  (Doing so might also suggest attractive modifications or generalizations.)
\item It would be desirable, in the spirit of unification, to combine the two algebras of  Eqns.\,(\ref{symmetryRelations}, \ref{qmCommutators}) into an integrated whole. 
\end{itemize}

Since it would be cruel of Nature to frustrate natural desires, and Nature is not cruel:

\begin{itemize}
\item {\it Quantization and fundamental symmetry will not appear as separate principles, but as two aspects of a deeper unity.}
\end{itemize}

\bigskip

\subsection{Unification VII: Mind and Matter}

\bigskip

Although many details remain to be elucidated, it seems fair to say that metabolism and reproduction, two of the most characteristic features of life, are now broadly understood at the molecular level, as physical processes.  Francis Crick's ``Astonishing Hypothesis'' is that it will be possible to bring understanding of basic psychology, including biological cognitive processing, memory, motivation, and emotion to a comparable level.    One might call this ``reduction'' of mind to matter.   

But mind is what it is, and what it is will not be diminished for being physically understood.   To me it seems both prettier and more sensible to regard the Astonishing Hypothesis anticipating the richness of behavior matter can exhibit, given what physics has told us matter is (well enough for this purpose).    I'd be thrilled to understand how I work -- what a trip!

\begin{itemize}
\item {\it Crick's ``Astonishing Hypothesis'' will continue to be valid and fruitful. Biological memory, cognitive processing, motivation, and emotion will be understood at the molecular level.}
\end{itemize}

And if physics evolves to describe matter in terms of information, as we discussed earlier, a circle of ideas will have closed.  Mind will have become more matter-like, and matter will have become more mind-like.

\bigskip

\section{Vistas}

\bigskip

\subsection{Making Things (Micro)}

Already in 1929 Dirac stated:
\begin{quote}
The underlying physical laws necessary for the mathematical theory of a large part of physics and the whole of chemistry are thus completely known, and the difficulty is only that the exact application of these laws leads to equations much too complicated to be soluble. It therefore becomes desirable that approximate practical methods of applying quantum mechanics should be developed, which can lead to an explanation of the main features of complex atomic systems without too much computation.
\end{quote}

Dirac's claim that we know the fundamental laws well enough to support calculation of any physical process relevant to chemistry, materials science, or anything else of practical interest looks better than ever.    On the other hand, the boundary defining what it is reasonable to interpret as ``much too complicated'' and  ``too much computation'' has changed radically.   Modern computing machines are many orders of magnitude more capable than anything available in 1929, and (as we shall discuss below) there are good reasons to think that within 100 years we will have many more orders of magnitude improvement.   

Are there materials that will support space elevators?  Are there room temperature superconductors?  These questions and any number of others, will become accessible as computers do for chemistry what they have already done for aircraft design, supplementing and ultimately supplanting laboratory experimentation with computation:

\begin{itemize}
\item {\it Calculation will increasingly replace experimentation in design of useful materials, catalysts, and drugs, leading to much greater efficiency and new opportunities for creativity.}
\end{itemize}

In recent years we have seen the beginnings of first-principles nuclear physics, here based on direct numerical solution of the equations of QCD and QED.  Very recently a major landmark was reached, as the neutron-proton mass difference was calculated with decent accuracy.

\begin{itemize}
\item {\it Calculation of many nuclear properties from fundamentals will reach $<$ 1\% accuracy, allowing  much more accurate modeling of supernovae and of neutron stars. Physicists will learn to manipulate atomic nuclei dexterously, as they now manipulate atoms, enabling (for example) ultra-dense energy storage and ultra-high energy lasers.}
\end{itemize}

\bigskip

\subsection{Making Things (Meso)}

Present-day mainstream computers are essentially two-dimensional. They are based on chips that must be produced under exacting clean room conditions, since any fault can be fatal to their operation.   And if they are damaged, they do not recover.   

Human brains differ in all those respects: they are three-dimensional, they are produced in messy, loosely controlled conditions, and they can work around faults or injuries.   There are strong incentives to achieve those features in systems that retain the density, speed, and scalability of semiconductor technology, and there is no clear barrier to doing so.   Thus:

\begin{itemize}
\item {\it Capable three-dimensional, fault-tolerant, self-repairing computers will be developed.  In engineering those features, we will learn lessons relevant to neurobiology.}
\end{itemize}

In a similar vein, we may aspire to make body-like machines as well as brain-like computers:

\begin{itemize}
\item{\it Self-assembling, self-reproducing, and autonomously creative machines will be developed.  Their design will adapt both ideas and physical modules from the biological world.}
\end{itemize}

\subsection{Making Things (Macro)}

Combining those ideas, we have:

\begin{itemize}
\item {\it Bootstrap engineering projects wherein machines, starting from crude raw materials, and with minimal human supervision, build other sophisticated machines -- notably including titanic computers -- will be underway. }
\end{itemize}

Freeman Dyson famously imagined ``Dyson spheres'', whereby all or most of the energy of a star would be captured by a surrounding shell or a cloud of interceptors, for use by an advanced technological civilization.   While that remains a distant prospect, the capture and use of a substantial part of solar energy impinging on Earth may be a necessity for human civilization, and seems eminently feasible to achieve within 100 years:

\begin{itemize}
\item {\it A substantial fraction of the Sun's energy impinging on Earth will be captured for human use.}
\end{itemize}

\subsection{Enhanced Sensoria}

Human perception leaves a lot on the table.   Consider, for example, color vision.   Whereas the electromagnetic signals arriving at our eyes contain a continuous range of frequencies, and also polarization, what we perceive as ``color'' is a crude hash encoding, where the power spectrum is lumped into three bins, and polarization is ignored.  (Compare our perception of sound, where we do a frequency analysis, and can appreciate distinct tones within chords.)   Also, of course, we are insensitive to frequencies outside the visible range, including ultraviolet and infrared.    Many other animals do finer sampling, so there is valuable information about our natural environment, not to mention possibilities for data visualization and art, to be gained by expanding color perception.   

Modern microelectronics and computing offer attractive possibilities for accessing this information.  By appropriate transformations, we can encode it in our existing channels, in a sort of induced synesthesia.  

\begin{itemize}
\item {\it We will vastly expand the human sensorium, opening the doors of perception.}
\end{itemize}

Physicists often, and rightly, admire the beauty of their concepts and equations.  On the other hand, humans are intensely visual creatures.  It will be fruitful to use modern resources of signal processing and computer graphics to translate the beautiful concepts and equations into forms such that physicists can bring their visual cortex fully to bear on them, and that people in general can admire and enjoy.   

\begin{itemize}
\item {\it Artists and scientists will work together, to create new works of extraordinary beauty.}
\end{itemize}

\bigskip

\subsection{Quantum Sensoria, Quantum Minds}

\bigskip

Quantum mechanics reveals other unseen infinities.   Perhaps the most characteristic quantum effect, which expands its state space exponentially, is entanglement.  But entanglement is both delicate and (since it involves correlations) subtle to observe, so our exploration of that central feature of the quantum world is only now properly commencing.   New kinds of observables will open to view, and new states of matter will be revealed: 

\begin{itemize}
\item {\it Measurement of entanglement, and measurement exploiting entanglement, will become major branches of physics.}
\end{itemize}

Quantum computing requires detailed control of entanglement, and the diagnostics of quantum computing will call on techniques to measure entanglement.  

Quantum computers could be of great service in living up to Dirac's challenge -- to me, a more inspiring prospect than factoring large numbers.  

\begin{itemize}
\item {\it Quantum computers supporting thousands of qubits will become real and useful.}
\end{itemize}

Artificial intelligence, in general, offers strange new possibilities for the life of mind.  An entity capable of accurately recording its state could purposefully enter loops, to re-live especially enjoyable episodes, for example.  Quantum artificial opens up possibilities for qualitatively new forms of consciousness.    A quantum mind could experience a superposition of ``mutually contradictory'' states, or allow different parts of its wave function to explore vastly different scenarios in parallel.  Being based on reversible computation, such a mind could revisit the past at will, and could be equipped to superpose past and present.   

\begin{itemize}
\item {\it Quantum intelligence will open up qualitatively new possibilities for the life of the mind.}
\end{itemize}

And who knows, maybe a quantum mind will even understand quantum mechanics!

\subsection{Changing Minds}

\subsubsection{Expanded Identity}

As people acquire routine access to extremely capable distant sensors and actuators, their sense of identity will expand beyond the limits of their physical bodies.   An immersive experience of ``being there'' will not necessarily involve being there, physically.   This will be an important element of the expansion of human culture beyond Earth, since human bodies are very poorly suited to extraterrestrial environments.   One can imagine an expanding web of intelligence more easily than an expanding web of settlement, and I expect it will happen much sooner.  The finite speed of information transfer (i.e., the speed of light) comes into play when different parts of an expanded identity communicate over astronomical distances; so people will need to, and will, come to internalize the subtleties of synchronization and time in special relativity.

\subsubsection{Wisdom}

\begin{itemize}

\item{\it Emergent Humility}

I have described a future in which people know much more about, and have vastly greater power over, the physical world than we do today.   Paradoxically, perhaps, I think that this will make them more sensitive to gaps in their knowledge, and ambitious to accomplish more.    

\begin{quote}
É to myself I seem to have been only like a boy playing on the sea-shore, and diverting myself in now and then finding a smoother pebble or a prettier shell than ordinary, whilst the great ocean of truth lay all undiscovered before me. -- Newton
\end{quote}

Such emergent humility reflects not so much modesty, as largeness of vision.

\item{\it Complementarity}

Complementarity -- the insight that there can be valid, but mutually exclusive ways of regarding a single reality -- is deeply embedded in the fabric of quantum theory.  As Niels Bohr recognized, it is a concept that applies much more broadly.  By being open to the possibility of understanding the world from different perspectives we can enrich our own experience of it -- and also learn tolerance.   

\end{itemize}

\pagebreak

\begin{center}
{\bf Appendix: Core Theory}
\end{center}

\bigskip

The quantum revolution gave this revelation: we've finally learned what Matter is.   The necessary equations are part of the theoretical structure often called the Standard Model.    That yawn-inducing name fails to convey the grandeur of the achievement, and does not serve fundamental physics well in its popular perception.   I'm going to continue my campaign, begun in {\it The Lightness of Being}, to replace it with something more appropriately awesome:
\begin{quotation}
{\small Standard Model} \ $\rightarrow$ \ {\large Core Theory}
\end{quotation}

This change is more than justified, because
\begin{enumerate}
\item ``Standard Model'' is used in several different ways.   Its original and appropriate use is to describe the electroweak $SU(2)\times U(1)$ gauge theory.   It is now often, but not always, taken to include QCD as well; and sometimes, but not always, taken to include general relativity as well.   Besides introducing ambiguity that packaging of self-contained, few-parameter theories together with (let alone within) a relatively messy container does not render QCD or general relativity justice .
\item ``Model'' connotes a disposable makeshift, awaiting replacement by the ``real thing''.  But the Core Theory is already an accurate representation of very large portions of physical reality, which any future, hypothetical ``real thing'' must take into account. 
\item ``Standard'' connotes ``conventional'', and hints at superior wisdom.  But no such superior wisdom is available.  In fact, I think -- and mountains of evidence attest -- that while the Core Theory will be supplemented, its core will persist.
\end{enumerate}

The Core Theory completes, for practical purposes, the analysis of matter.  Using it, we can {\it deduce\/} what sorts of atomic nuclei, atoms, molecules -- and stars -- exist.  And we can reliably orchestrate the behavior of larger assemblies of these elements, to make transistors, lasers, or Large Hadron Colliders.    
The equations of the Core Theory have been tested with far greater accuracy, and under far more extreme conditions, than are required for applications in chemistry, biology, engineering, or astrophysics.    While there certainly are many things we don't understand, we do understand the Matter we're made from, and that we encounter in normal life -- even if we're chemists, engineers, or astrophysicists.

\bigskip

\end{document}